\pdfoutput=1
\documentclass[a4paper]{article}
\usepackage[a4paper,margin=2.5cm]{geometry}

\usepackage{graphicx}
\usepackage[output-product=\cdot,exponent-product=\cdot,per-mode=fraction,detect-weight=true]{siunitx}
\usepackage{amsmath}
\usepackage{amsfonts}
\usepackage{amssymb}
\usepackage{mathtools}
\usepackage{bm}
\usepackage{commath}
\usepackage{booktabs}
\usepackage{tabulary}
\usepackage{placeins}
\usepackage{enumitem}
	\newlist{step}{enumerate}{1}
	\setlist[step]{label=Step~\arabic*:,leftmargin=*}
	\newlist{result}{enumerate}{1}
	\setlist[result]{label=R\arabic*:,leftmargin=*}
\usepackage{marvosym}
\usepackage[raggedright]{titlesec}

\newcommand\blfootnote[1]{%
  \begingroup
  \renewcommand\thefootnote{}\footnote{#1}%
  \addtocounter{footnote}{-1}%
  \endgroup
}

\usepackage[
	  backend=biber
	, style=authoryear
	, maxcitenames=2
	, maxbibnames=99
	, giveninits=true
	, uniquename=init
	, bibencoding=utf8
	, natbib=true
	, urldate=edtf,date=edtf,seconds=true
	  ]{biblatex}
	\bibliography{refs}
	\AtEveryBibitem{%
		\clearfield{doi}
	}

\usepackage[hidelinks]{hyperref}

\newcommand{\astar}{\texorpdfstring{A$^\star$}{A*}}

\newcommand{\tr}{^\top}
\newcommand{\mtrx}[1]{\bm{#1}}
\newcommand{\vect}[1]{\bm{#1}}
\newcommand{\mset}[1]{\mathbb{#1}}
\newcommand{\real}{\mathbb{R}}

\DeclareMathOperator{\scirc}{S}
\title{Improvements to Warm-Started Optimized Trajectory Planning for ASVs}
\author{Glenn Bitar~\Letter, Anastasios M.\ Lekkas and Morten Breivik}
\date{}

\begin{document}
	\maketitle
	\begin{abstract}

We present improvements to a recently developed method for trajectory planning for autonomous surface vehicles (ASVs) in terms of run time.
The original method combines two types of planners: An \astar{} implementation that quickly finds the global shortest piecewise linear path on a uniformly discretized map, and an optimal control-based trajectory planner which takes into account ASV dynamics.
Firstly, we propose an improvement to the discretization of the map by switching to a Voronoi diagram rather than the uniform discretization, which offers a far more sparse search tree for the \astar{} implementation.
Secondly, modifications to the path refinement are made, as suggested in a paper by \citeauthor{Bhattacharya_2008}.
The changes result in a reduction to the run time of the first part of the method of \SI{85}{\percent} for an example scenario while maintaining the same level of optimality.

	\end{abstract}
	\blfootnote{%
		The authors are with the Centre for Autonomous Marine Operations and Systems, and with the Department of Engineering Cybernetics, Norwegian University of Science and Technology (NTNU), NO-7491 Trondheim, Norway.
		E-mails: \{glenn.bitar,anastasios.lekkas\}@ntnu.no, morten.breivik@ieee.org.
		The work is funded by the Research Council of Norway and Innovation Norway with project number 269116.
		Additionally, the work is supported by the Centres of Excellence funding scheme with project number 223254.
	}
	\FloatBarrier

\section{Introduction} % (fold)
\label{sec:introduction}

Development of technology for autonomous surface vehicles (ASVs) is continuing at a rapid pace, motivated by factors such as economy, safety and the environment.
In addition to academic interest, commercial organizations are diving into the use of autonomous technology.
Among other use cases ASVs have been instrumental to reducing costs for ocean floor mapping surveys, for instance in Alaska in 2016 \citep{Orthmann2016AsvMapping}.

At the base of any autonomous ship operation is a motion planning and control system.
This type of system is responsible for planning and executing safe motion trajectories that avoid collision, and that promote some objective, such as minimum time or energy consumption.

\begin{figure}[tb]
	\centering
	\includegraphics[width=0.75\linewidth]{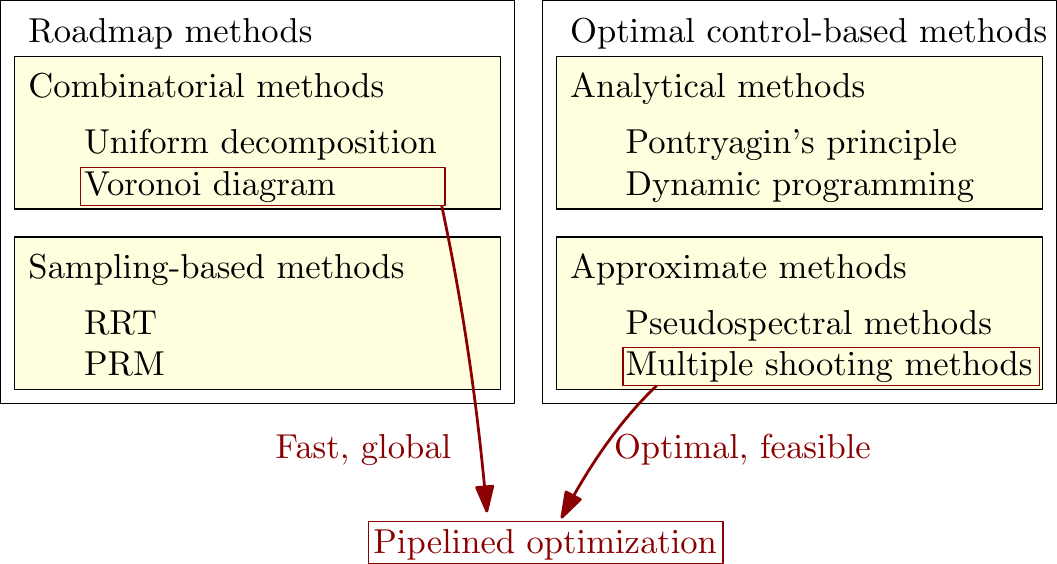}
	\caption{%
		Categorization of some motion planning algorithms.
	}
	\label{fig:motion-planning-categorization}
\end{figure}

Several types of motion planners have been researched and developed for marine applications.
In \autoref{fig:motion-planning-categorization} we have classified some types of motion planning methods into \emph{roadmap methods} and \emph{optimal control-based methods}.
Roadmap methods build a path in the free space by exploring discrete points on the map.
Further, the roadmap methods can be divided into combinatorial methods that discretize all of the free space, and sampling-based methods that randomly explore points on the map.
Optimal control-based methods build upon optimization, and are divided into analytical methods that usually can only find solutions for very simple scenarios, and approximate methods which are more practical in real-world scenarios.
A more in-depth background on these motion planning classes is provided in \citep{Bitar2019CamsPipeline}.

A combinatorial roadmap approach that is commonly used in path planning algorithms is the Voronoi diagram \citep{Aurenhammer1991Voronoi}.
The Voronoi diagram splits the workspace in cells according to generator points, and is used in e.g.\ aerial \citep{Bortoff_2000} and marine \citep{Candeloro_2017,Gold_2016_tesselations} applications.
\citet{Gold_2016_tesselations} shows how Voronoi diagrams may be used to build and dynamically update 2D and 3D structures in geographic information systems.
\citet{Bortoff_2000} proposes a method where the path generated from the Voronoi diagram is used as an initial condition for a set of differential equations which at steady state optimize a flight path for a combination of distance to enemy radars and path length.
The Voronoi diagram is also used as part of a more complex path planning scheme by \citet{Candeloro_2017}.
Here the Voronoi diagram provides an initial path, which is subsequently refined, and then smoothed using Fermat's spiral segments.
Despite the practical nature of such solutions, they cannot guarantee that certain objectives, outside the shape of the path itself, will be satisfied.

In \citep{Bitar2019CamsPipeline} we have developed a pipelined trajectory planning method that combines advantages from a roadmap method and optimal control.
That method builds a trajectory from a start point to a goal point in a map of obstacles through three steps:
\begin{step}
	\item Discretize the obstacle map into a uniform grid, and perform an \astar{} search to give the shortest path.
	\item Refine path and add artificial temporal information to convert it into a trajectory.
	\item Solve an optimal control problem (OCP) which gives an optimized trajectory from start to goal, and warm start the solver with information from Step~2.
\end{step}
This paper presents improvements to \citep{Bitar2019CamsPipeline} in terms of run time through the following contributions:
\begin{itemize}
	\item The uniform grid that discretizes the map in Step~1 of \citep{Bitar2019CamsPipeline} is replaced with a Voronoi diagram, which requires far fewer discrete nodes, increasing search speed.
	\item The path refinement process in Step~2 is improved with ideas from \citep{Bhattacharya_2008}.
	\item The integration process in Step~2 that creates part of the initial guess for the OCP solver is modified to increase speed.
\end{itemize}
The improvements cause a significant reduction in the run time of steps~1 and~2, while maintaining the same level of optimality.

Preliminary topics are presented in \autoref{sec:preliminaries}.
In \autoref{sec:trajectory_planning_method} we briefly present the pipelined method from \citep{Bitar2019CamsPipeline}, and describe the improved method in more detail.
Simulation scenarios and results are presented in \autoref{sec:simulation_scenarios_and_results}, and we conclude the paper in \autoref{sec:conclusion}.

% section introduction (end)

\section{Preliminaries} % (fold)
\label{sec:preliminaries}

In this paper we utilize several tools that will be briefly explained in this section.

\subsection{Voronoi diagram} % (fold)
\label{sub:voronoi_diagram}

A Voronoi diagram \citep{Aurenhammer1991Voronoi} partitions a plane into regions based on distance to \emph{Voronoi generator points}, or just generators for brevity.
These regions consist of all points closer to their generator than to any other generator.
We refer to those regions as \emph{Voronoi cells}.
A Voronoi diagram is useful in path planning since the edges of the Voronoi cells are distant to all obstacle in a map, if the generators lie along the obstacle boundaries.
The Voronoi cell edges are then candidates for paths in the map, and connect \emph{Voronoi vertices} in a searchable graph.

A distinct advantage of using Voronoi diagrams in path planning is that the graph can contain all possible paths between obstacles, while the map is sparsely discretized (i.e.\ with a low number of Voronoi vertices).
An inherent disadvantage is that the Voronoi edges will be placed far away from the obstacle boundaries, which increases path lengths.
This disadvantage is addressed by using contributions from \citep{Bhattacharya_2008}.

% subsection voronoi_diagram (end)

\subsection{\astar{} search} % (fold)
\label{sub:astar}

To be able to find an optimal (e.g.\ shortest) path in a map with the help of a Voronoi diagram, it is necessary to search the graph generated by the Voronoi diagram in some way.
In this paper we use \astar{} to search for the shortest path \citep{Hart1968}.
\astar{} is a graph search algorithm guided by a heuristic which quickly leads the search to the desired node.

% subsection astar (end)

\subsection{ASV modeling} % (fold)
\label{sub:asv_modeling}

We use a model-based approach when generating our optimized trajectory.
The ASV is modeled mathematically by a set of differential equations which we will present here.
The equations are retrieved from \citep{Bitar2019CamsPipeline}.

A simple nonlinear 3-degree-of-freedom ship model is used, and the model has the form
\begin{subequations}
	\label{eq:ship-model}
	\begin{align}
		\dot{\vect{\eta}} &= \mtrx{R}(\psi) \vect{\nu}\,, \\
		\mtrx{M} \dot{\vect{\nu}} + \mtrx{C}(\vect{\nu}) \vect{\nu} + \mtrx{D}(\vect{\nu}) \vect{\nu} &= \vect{\tau}(\vect{u})\,.
	\end{align}
\end{subequations}
The pose vector $\vect{\eta} = [x, y, \psi]\tr \in \real^2 \times \scirc$ contains the ASV's position and heading angle in the Earth-fixed North-East-Down (NED) frame.
The velocity vector $\vect{\nu} = [u, v, r]\tr \in \mset{R}^3$ contains the ASV's body-fixed velocities: surge, sway and yaw rate, respectively.
The rotation matrix $\mtrx{R}(\psi)$ transforms the body-fixed velocities to NED:
\begin{equation}
	\mtrx{R}(\psi) =
	\begin{bmatrix}
		\cos \psi & -\sin \psi & 0 \\
		\sin \psi & \cos \psi & 0 \\
		0 & 0 & 1
	\end{bmatrix}.
\end{equation}
The matrix $\mtrx{M} \in \real^{3\times3}$ represents system inertia, $\mtrx{C}(\vect{\nu}) \in \real^{3\times3}$ Coriolis and centripetal effects, and $\mtrx{D}(\vect{\nu}) \in \real^{3\times3}$ represents damping effects.
The ASV is controlled by the control vector $\vect{u} = [X, N]\tr \in \real^2$, which contains surge force and yaw moment.
The control vector is mapped to a force vector $\vect{\tau}(\vect{u}) = [X, 0, N]\tr$.

% subsection asv (end)

\subsection{Optimal control problem} % (fold)
\label{sub:optimal_control_problem}

To find an optimized trajectory, we solve an OCP, which is the challenge of finding a control and state trajectory that minimizes an objective functional, while satisfying both differential and algebraic constraints.
In our case, we pose the trajectory planning problem as an OCP in the following form, retrieved from \citep{Bitar2019CamsPipeline}:
\begin{subequations}
\label{eq:ocp}
\begin{align}
	\label{eq:ocp-cost}
	&\min_{\vect{\eta}(\cdot), \vect{\nu}(\cdot), \vect{u}(\cdot)} \int_{0}^{t_{\text{max}}} F(\vect{\eta}(t), \vect{\nu}(t), \vect{u}(t)) \dif t \\
	\nonumber
	&\text{subject to} \\
	\label{eq:ocp-dynamics-eta}
	&\dot{\vect{\eta}} = \mtrx{R}(\psi) \vect{\nu} ~ \forall t \in [0, t_{\text{max}}], \\
	\label{eq:ocp-dynamics-nu}
	&\mtrx{M} \dot{\vect{\nu}} + \mtrx{C}(\vect{\nu}) \vect{\nu} + \mtrx{D}(\vect{\nu}) \vect{\nu} = \vect{\tau}(\vect{u}) ~ \forall t \in [0, t_{\text{max}}], \\
	\label{eq:ocp-ineq}
	&\vect{h}(\vect{\eta}(t), \vect{\nu}(t), \vect{u}(t)) \leq \vect{0} ~ \forall t \in [0, t_{\text{max}}], \\
	\label{eq:ocp-boundary}
	&\vect{e}(\vect{\eta}(0), \vect{\nu}(0), \vect{\eta}(t_{\text{max}}), \vect{\nu}(t_{\text{max}})) = \vect{0}\,.
\end{align}
\end{subequations}

The solution of the OCP gives state and input trajectories that optimize the cost functional \eqref{eq:ocp-cost}, consisting of the cost-to-go function
\begin{equation}
	\label{eq:cost-to-go}
	F(\vect{\eta}, \vect{\nu}, \vect{u}) = K_e F_e(\vect{\nu}, \vect{u}) + K_t F_t(\vect{\nu})\,,
\end{equation}
with tuning parameters $K_e, K_t > 0$.
The first term penalizes energy usage and describes work done by the actuators, while the second term is a disproportionate penalization on turn-rate $r$, and prefers readily observable turns performed with high turn-rate.
The idea for the turn-rate penalization is obtained from \citep{Eriksen2017MPC}.
The same cost-to-go function is used in \citep{Bitar2019CamsPipeline}.

The solution will satisfy the dynamic constraints \eqref{eq:ocp-dynamics-eta} and \eqref{eq:ocp-dynamics-nu}, as well as the inequality constraint \eqref{eq:ocp-ineq} which encodes obstacles and state constraints, and \eqref{eq:ocp-boundary}, which are the start and goal conditions.

% subsection optimal_control_problem (end)

\subsection{Transcription and solver} % (fold)
\label{sub:transcription_and_solver}

In order to solve the OCP from \autoref{sub:optimal_control_problem}, it is transcribed to a nonlinear program (NLP) by using a multiple shooting approach.
The transcription gives the following NLP:
\begin{subequations}
\label{eq:nlp}
\begin{align}
	\label{eq:nlp-cost}
	&\min_{\vect{w}} \phi(\vect{w}) \\
	\nonumber
	&\text{subject to} \\
	\label{eq:nlp-ineq}
	& \vect{g}_{lb} \leq \vect{g}(\vect{w}) \leq \vect{g}_{ub}, \\
	\label{eq:nlp-bounds}
	&\vect{w}_{lb} \leq \vect{w} \leq \vect{w}_{ub} \,.
\end{align}
\end{subequations}
The details of the NLP structure are found in \citep{Bitar2019CamsPipeline}.

The NLP \eqref{eq:nlp} is non-convex, and a good solution is hard to find without a ``good'' initial guess.
Solvers generally find a local optimum, and providing the solver with an initial guess is referred to as warm-starting the solver.
To solve the NLP we use Casadi \citep{Andersson2018casadi} for Matlab, with the Ipopt solver \citep{Wachter2005ipopt}.

% subsection transcription_and_solver (end)

% section preliminaries (end)

\section{Trajectory planning method with improvements} % (fold)
\label{sec:trajectory_planning_method}

\begin{figure}[tb]
	\centering
	\includegraphics[width=0.8\linewidth]{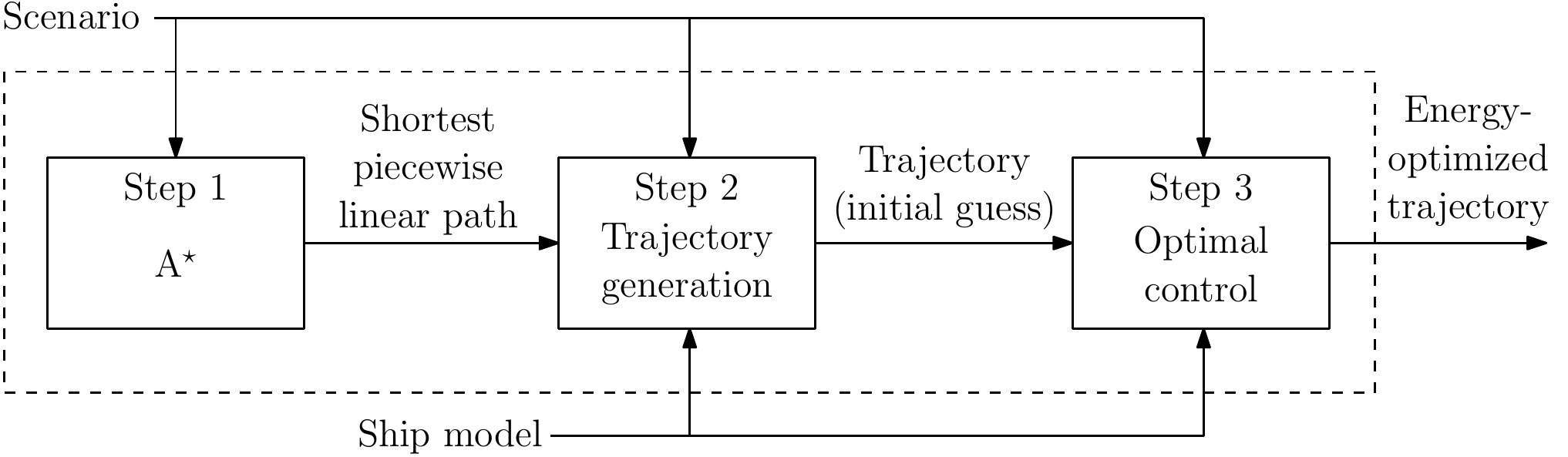}
	\caption{%
		Pipelined concept.
	}
	\label{fig:concept-block-diagram}
\end{figure}

\begin{table}[tb]
	\centering
	\caption{Pipelined trajectory planning method steps.}
	\begin{tabulary}{1.0\linewidth}{RLLL}
		\toprule
		Step & Original method \citep{Bitar2019CamsPipeline} & Improved method & Qualitative changes \\
		\midrule
		1 &		Discretize map into a uniform grid with cell sizes $\Delta d \times \Delta d$. Search from start to goal with \astar{}. & 
				Discretize map into a Voronoi diagram using generators with spacing $\Delta d$ on obstacle boundaries. Search from start to goal with \astar{}. &
				Improvement results in a far more sparse search tree that still contains paths between all obstacles, resulting in a faster \astar{} search. \\[0.2cm]
		2 & 	Perform waypoint reduction with a custom algorithm. Connect the path with circle arcs, and add artificial temporal information. Use Runge-Kutta 4 integration for cost-to-go propagation. &
				Perform waypoint reduction and path refinement using algorithms from \citep{Bhattacharya_2008}. Connect the path with circle arcs, and add artificial temporal information. Use improved Euler for cost-to-go propagation. &
				Improvement results in a good and close-to-feasible initial guess for Step~3. The improved Euler propagation is much faster than Runge-Kutta 4, since there is no need for interpolation. \\[0.2cm]
		3 & 	Solve NLP \eqref{eq:nlp} to retrieve a solution of the OCP \eqref{eq:ocp}, with the trajectory from Step~2 as an initial guess for warm starting. &
				Same as in \citep{Bitar2019CamsPipeline}. &
				No changes. \\
		\bottomrule
	\end{tabulary}
	\label{tab:method_steps}
\end{table}

In \citep{Bitar2019CamsPipeline} we have developed a pipelined three-step method to obtain a dynamically feasible and optimized trajectory between a start and goal position in a map.
Figure~\ref{fig:concept-block-diagram} shows how the pipelined concept works.
In this paper, improvements are made to the first two steps in terms of run time, and a comparative table of the steps in the method is provided in \autoref{tab:method_steps}.
The uniform grid discretization of Step~1 in \citep{Bitar2019CamsPipeline} is advantageous in the sense that an arbitrary level of precision can be obtained by reducing the grid size $\Delta d$.
However, this leads to a large and dense search space, which results in a long run time of the \astar{} algorithm.
As described in \autoref{sub:voronoi_diagram}, a Voronoi discretization of the map gives a sparse search tree, while still containing all possible routes.
For this reason, we switch to the Voronoi discretization.
The Voronoi generators are points along the boundary of each obstacle, as well as along the map edge.
They are spaced with a distance of $\Delta d$, which is replacing the meaning of $\Delta d$ in \citep{Bitar2019CamsPipeline}.

The disadvantage of using a Voronoi discretization is also described in \autoref{sub:voronoi_diagram};
it leads to longer paths, since the Voronoi cell edges are inherently placed distant from each obstacle.
This motivates the adding of a path refinement process to Step~2 of the method in order to shorten the path.
The path refinement process is the one developed by \citet{Bhattacharya_2008}, and consists of iteratively cutting corners of two edges by trying to make collision-free connections between intermediate points on the edges.
We also replace the waypoint reduction process from \citep{Bitar2019CamsPipeline} with the more efficient one described in \citep{Bhattacharya_2008}.

An additional improvement to the run time of Step~2 from \citep{Bitar2019CamsPipeline} is achieved by changing the integration process for obtaining the warm-started guess of the cost functional \eqref{eq:ocp-cost}.
Instead of performing Runge-Kutta 4 integration of \eqref{eq:cost-to-go}, we use the improved Euler method, removing the need for interpolation of the warm-starting state and input trajectories.

% section trajectory_planning_method (end)

\section{Simulation scenarios and results} % (fold)
\label{sec:simulation_scenarios_and_results}

We test the improved method on the example scenario presented in \citep{Bitar2019CamsPipeline}, a small island group north of Stavanger in Norway.
First, we present the difference in the discretization in Step~1 of the two results.
To demonstrate the advantage of using a Voronoi diagram discretization, we also include a version of the original method \citep{Bitar2019CamsPipeline} where the uniform grid size $\Delta d$ is significantly increased.
That results in a discretization that contains the same number of nodes as the Voronoi diagram discretization in our example.
For convenient referencing, we enumerate the three different results as:
\begin{result}
	\item The improved method with Voronoi discretization. Voronoi generator spacing of $\Delta d = \SI{100}{\meter}$. The discretized grid contains 833 non-colliding nodes.
	\item The original method with dense discretization, as used in \citep{Bitar2019CamsPipeline}. Grid size of $\Delta d = \SI{50}{\meter}$. The discretized grid contains 22929 non-colliding nodes.
	\item The original method with sparse discretization. Grid size of $\Delta d = \SI{270}{\meter}$. The discretized grid contains 843 non-colliding nodes.
\end{result}

\begin{figure}[tb]
	\centering
	\includegraphics[width=0.75\linewidth]{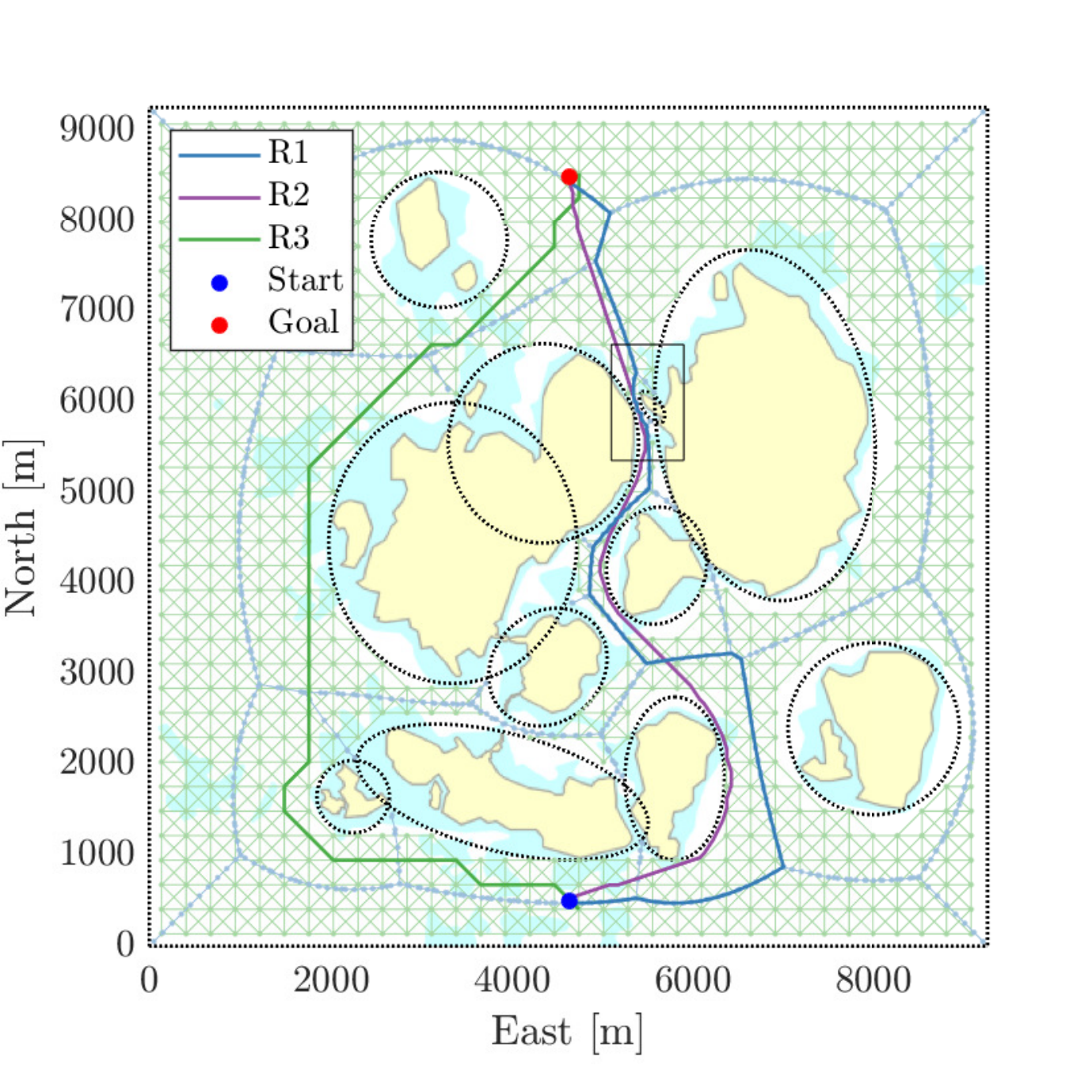}
	\caption{%
		Grids and shortest piecewise linear paths from Step~1 of R1 and R3.
		The grid from R2 is not shown due to its high density, but is shown in a zoomed-in area in \autoref{fig:scenario_1_grids_zoomed}.
	}
	\label{fig:scenario_1_grids}
\end{figure}

\begin{figure}[tb]
	\centering
	\includegraphics[width=0.75\linewidth]{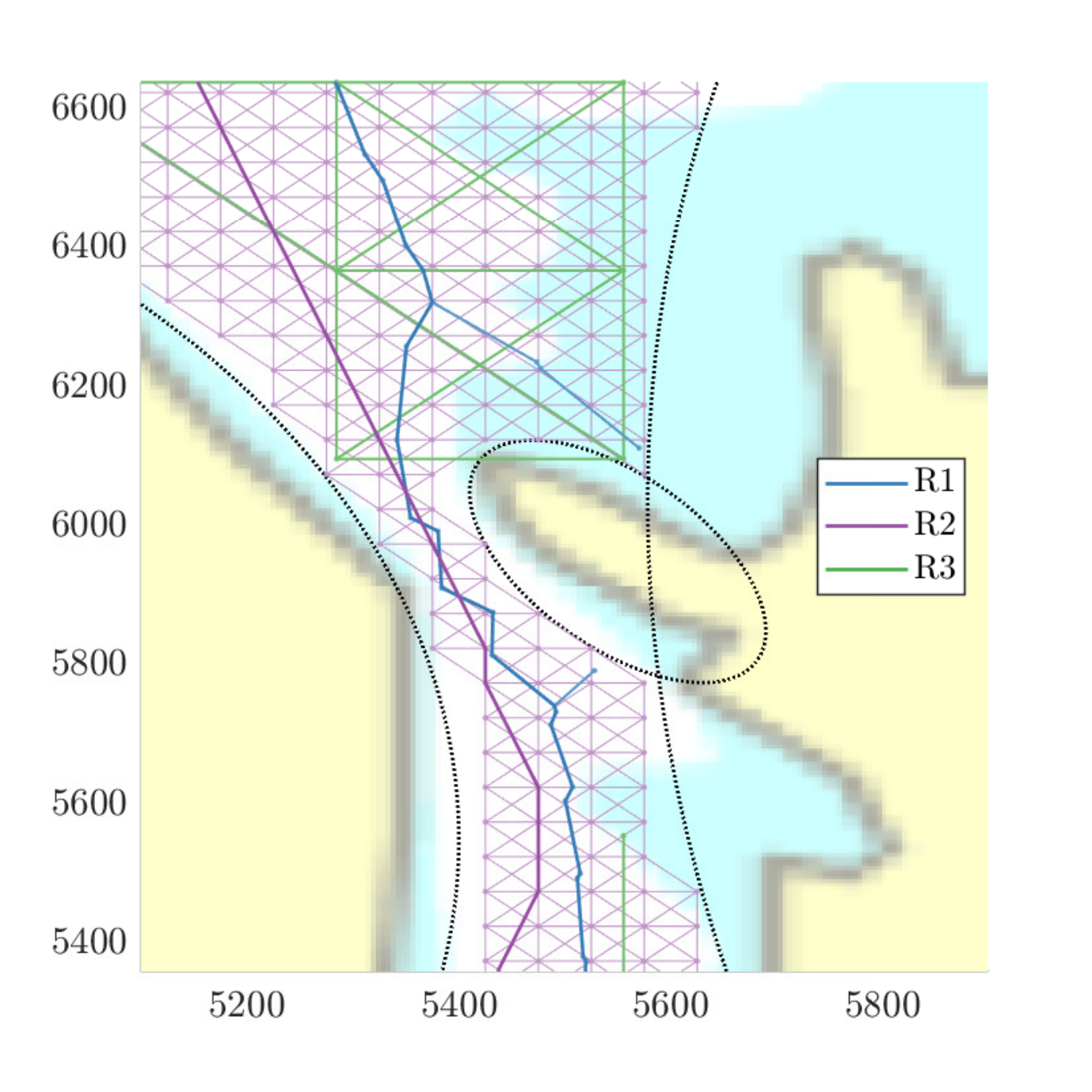}
	\caption{%
		A zoomed-in version of \autoref{fig:scenario_1_grids} with the grids from R1, R2 and R3.
	}
	\label{fig:scenario_1_grids_zoomed}
\end{figure}

Figures~\ref{fig:scenario_1_grids} and~\ref{fig:scenario_1_grids_zoomed} show the discretized grids of the three results.
We see that the Voronoi diagram proposes all possible routes between the obstacles as paths.
The dense grid from R2 also has all possible routes included in the search space, and the shortest piecewise linear path from those two grids go through the narrow passage in the central part of the map.
R3 has a much sparser discretization than R2, which causes the grid to be disconnected in the central narrow passage, leaving it out of the search space.
This causes the search to find a different route in R3 than the routes from R1 and R2, to the left in the map.

\begin{figure}[tb]
	\centering
	\includegraphics[width=0.5\linewidth]{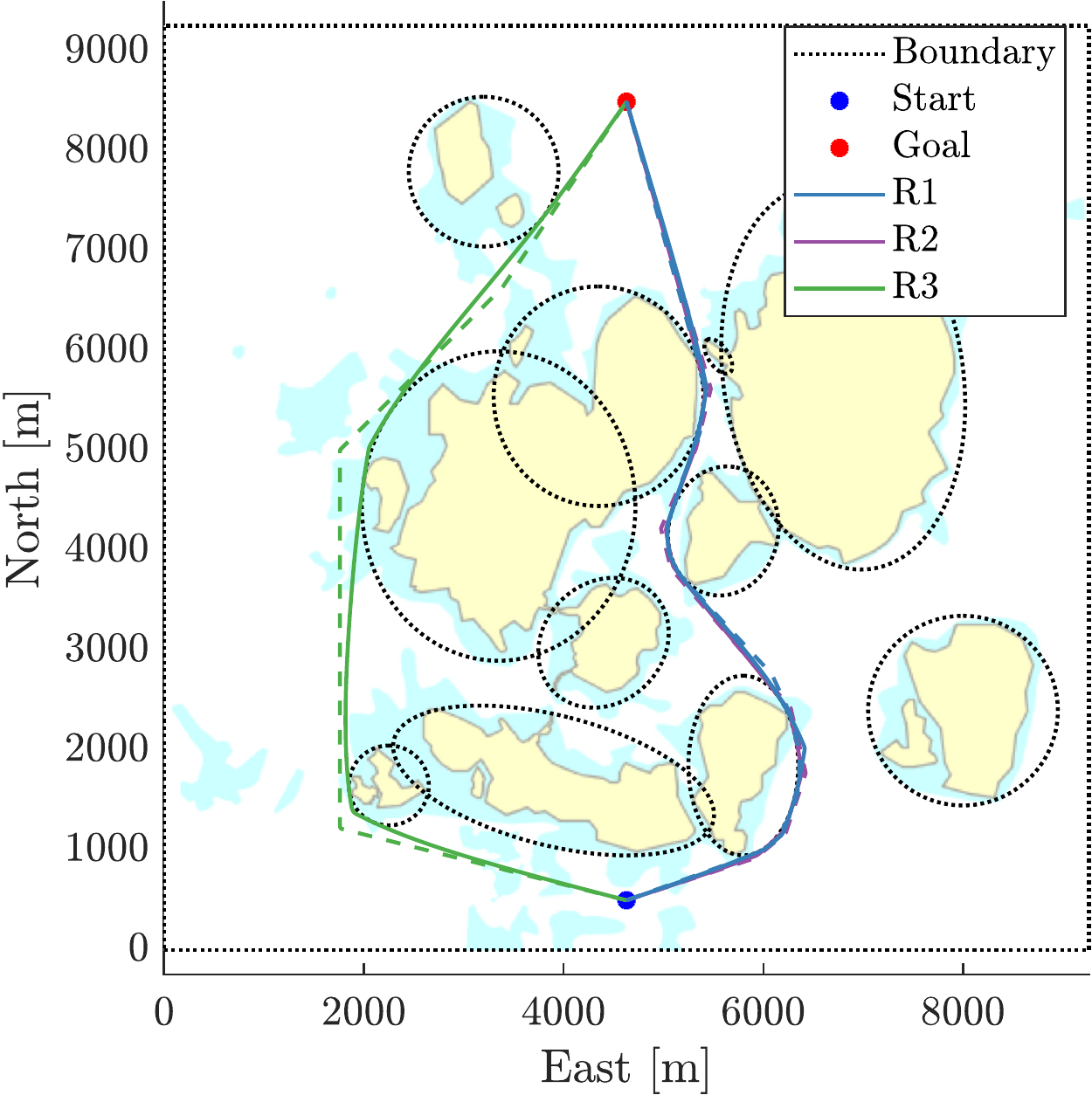}
	\caption{%
		Map and resulting trajectories for R1, R2 and R3.
		Warm-starting trajectories from Step~2 are shown in dashed lines.
	}
	\label{fig:scenario_1}
\end{figure}

Figure~\ref{fig:scenario_1} shows the trajectories from the three results, along with their warm-start guesses.
We have used the same parameters as in \citep{Bitar2019CamsPipeline}.
Since the shortest piecewise linear paths from Step~1 in both R1 and R2 go through the central narrow passage, the OCP from Step~3 finds similar locally optimal trajectories.
On the other hand, Step~1 in R3 finds a path to the left in the map, causing a longer trajectory.

\begin{table}[tb]
	\centering
	\caption{Results.}
	\begin{tabulary}{1.0\linewidth}{Lrrr}
		\toprule
				& R1 & R2 & R3 \\
		\midrule
		Energy cost 			& \SI{2.75e7}{[\joule]}			& \SI{2.74e7}{[\joule]} & \SI{3.85e7}{[\joule]} \\
		Run time 				& \SI{17.4}{[\second]}			& \SI{26.7}{[\second]} 	& \SI{18.5}{[\second]} \\
		\hspace{0.5cm} Step~1 	& \textbf{0.7}\,\si{[\second]}	& \SI{3.4}{[\second]} 	& \textbf{0.3}\,\si{[\second]} \\
		\hspace{0.5cm} Step~2 	& \textbf{0.1}\,\si{[\second]}	& \SI{2.2}{[\second]} 	& \SI{2.3}{[\second]} \\
		\hspace{0.5cm} Step~3 	& \SI{16.6}{[\second]}			& \SI{21.1}{[\second]} 	& \SI{15.9}{[\second]} \\
		Step~3 iterations 		& 72 							& 58 					& 43 \\
		\bottomrule
	\end{tabulary}
	\label{tab:scenario_1_results}
\end{table}

\autoref{tab:scenario_1_results} shows how R1 and R2 give virtually equal energy consumption, but with a large reduction in run times for steps~1 and~2 in R1.
The reduction in run time of Step~1 from R2 to R1 is attributed to the large reduction of nodes, simplifying the \astar{} search.
The reduction in run time of Step~2 in the same comparison is attributed to the improved waypoint reduction method, and the change of integration method, as detailed in \autoref{sec:trajectory_planning_method}.
Comparing the run times of Step~1 in R1 and R3, we see that it is faster in R3.
This may be attributed to the increased complexity in performing the discretization with a Voronoi diagram, compared to the uniform grid.
For Step~3, the Ipopt solver uses more steps for R1 than R2, but also shorter time.
These variations are due to the difference in initial guess.
However, we cannot say that this is a consistent effect from using the modified algorithms in steps~1 and~2, but state that they are arbitrary effects.

\begin{figure}[tb]
	\centering
	\includegraphics[width=0.5\linewidth]{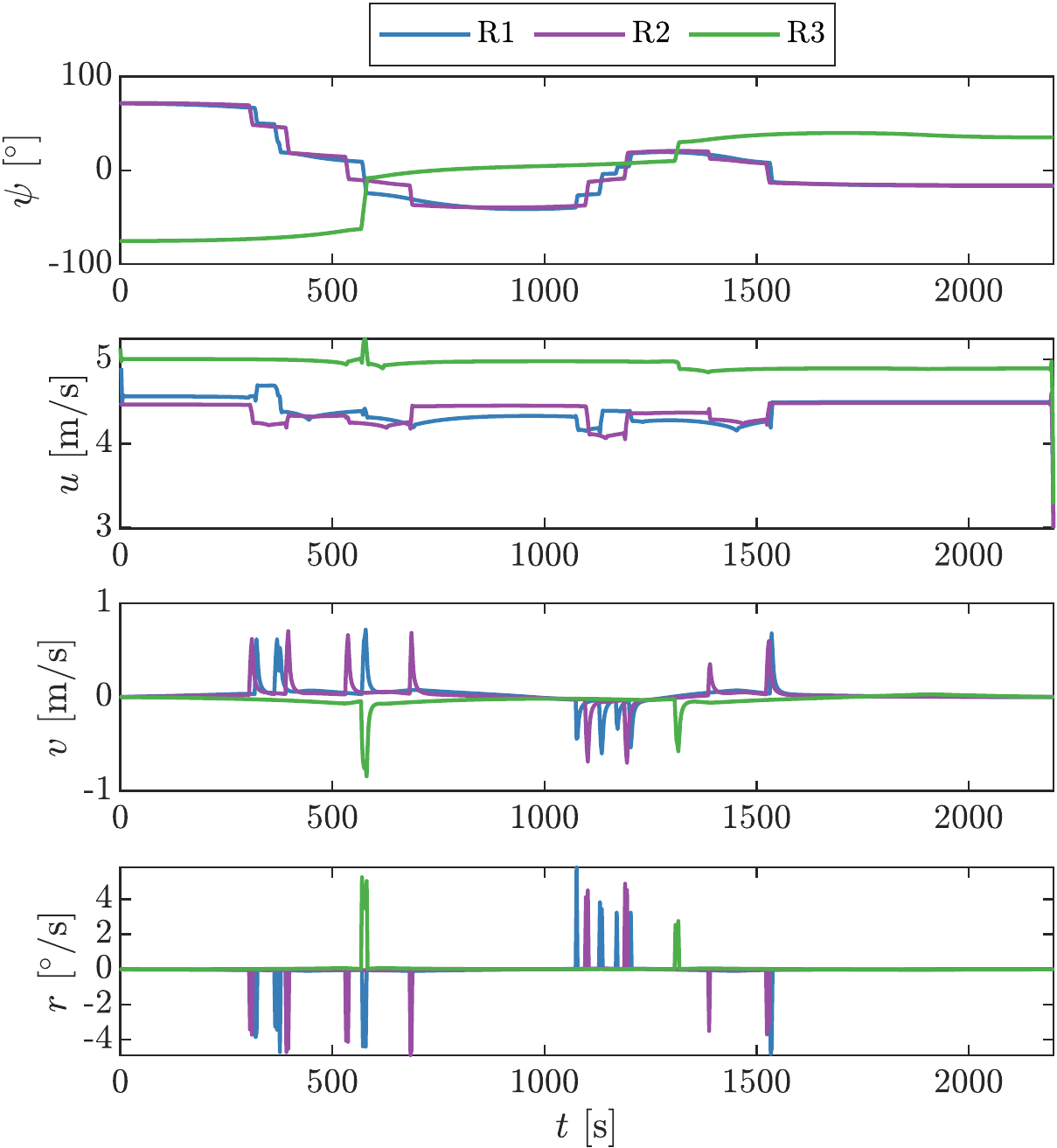}
	\caption{Heading, surge and sway velocity and yaw rate trajectories for R1, R2 and R3.}
	\label{fig:scenario_1_states}
\end{figure}

Figure~\ref{fig:scenario_1_states} shows the state trajectories for the three results.
It is worth noting the higher average surge velocity $u$ in R3 compared to the other two results.
This is of course due to the increased length of the trajectory, with the same amount of time to complete it.

% section simulation_scenarioes_and_results (end)

\section{Conclusion} % (fold)
\label{sec:conclusion}

We have presented improvements in the pipelined trajectory planning algorithm developed in \citep{Bitar2019CamsPipeline}.
The first improvement is to Step~1 of the algorithm, namely the discretization of the map in which to search for a global shortest path to goal.
The uniform grid used in \citep{Bitar2019CamsPipeline} is changed to one based on Voronoi diagrams, which results in a sparse discretization of the search space that still contains paths between all obstacles.
Run time for Step~1 is reduced from \SI{3.4}{\second} to \SI{0.7}{\second} in an example scenario.
Because of the longer paths resulting from the use of Voronoi diagrams, a path refinement process is added to Step~2, which shortens the path.
Changes to the waypoint reduction algorithm, and to the integration scheme used for finding an initial guess for the cost functional in Step~2, result in a reduction in Step~2 run time from \SI{2.2}{\second} to \SI{0.1}{\second}.
Together this results in a reduction of \SI{85}{\percent} in run time for steps~1 and~2 in the comparison scenario.

As with the original method from \citep{Bitar2019CamsPipeline}, the method is complete in terms of the shortest path, since this is the objective of the global \astar{} search.
The Voronoi diagram keeps paths between all obstacles in the search space even when they are coarsely discretized.
The implemented OCP provides local optimality in the sense of a given objective function.
Alone, the OCP solver will not find a global optimum, but guided by the initial guess, the OCP solver finds a local minimum which may be close to the global optimum, depending on the objective function.
The OCP will also return a feasible trajectory, as the ASV dynamics are taken into account.

% section conclusion (end)

	\FloatBarrier
	\printbibliography
\end{document}